\documentclass[aps,prb,twocolumn,superscriptaddress,showpacs]{revtex4}
\usepackage{ae}
\usepackage[T1]{fontenc}
\usepackage[ansinew]{inputenc}
\usepackage{amsmath}
\usepackage{amssymb}
\usepackage{graphicx}
\usepackage{color}
\usepackage[colorlinks]{hyperref}
\usepackage{epstopdf}

\begin{document}


\title{Frequency dependent dielectric response of ferroelectric-dielectric junction with negative electric capacitance}

\author{M. Piska\v{c}}
\author{D. Radi\'{c}*}
\affiliation{Department of Physics, Faculty of Science, University of Zagreb, Bijeni\v{c}ka 32, Zagreb 10000, Croatia}

\begin{abstract}
We calculated the frequency dependent dielectric response (electric susceptibility) of layered ferroelectric-dielectric junction, biased by the time-dependent harmonic voltage with single frequency $\omega$. Working point is stabilized, by the charge boundary condition between the layers, in the region with negative electric capacitance. The static susceptibility $\chi_0$ is negative and relative dielectric constant $\epsilon_r$ smaller than one, clearly indicating the opposite direction of electric field and polarization in the ferroelectric layer due to the negative electric capacitance. At finite frequencies this sign is preserved in real part of susceptibility which gains the frequency dependence. Also, frequency dependent imaginary part arises due to the phase shift between electric field and polarization. The type of that frequency dependence in linear regime is so-called relaxation (Debye) response, \textit{i.e.} $\chi'(\omega)=\chi_0/(1+(\tau \omega)^2)$ and $\chi"(\omega)=\chi_0\tau \omega/(1+(\tau \omega)^2)$, where $\tau$ is polarization switching time characteristic to ferroelectric material. In particular, we modeled the junction of ferroelectric BaTiO$_3$ and dielectric Al$_2$O$_3$, taking the experimental values of material parameters, and addressed the role of nonlinearity with respect to result of the linear response theory.
\end{abstract}

\maketitle

\bigskip
\textbf{I. Introduction}
\bigskip

The concept of negative electric capacitance, and especially experimental realizations of heterostructures featuring this property, have been an intriguing topic of research for over a decade.
The negative electric capacitance in ferroelectric material refers to the presence of local maximum of free energy, depending on polarization, around which the Taylor expansion effectively yields square of charge divided by double capacitance which has to be negative since we have the upside-down parabola (analogous to effective mass of the hole in the vicinity of the top of electron band). The problem is that this solution is unstable and it is never realized in the stand-alone ferroelectric in which the polarization settles in one of the stable minima surrounding the mentioned local maximum.
One motivation to deal with negative electric capacitance, as suggested by Salahuddin et al. \cite{Salahuddin}, is related to lower the subthreshold slope in the field effect transistors to less than standard 60 mV of channel potential per decade in current by replacing the standard insulator between the gate and channel with a ferroelectric layer which serves as an amplifier of a gate voltage. One advantage of integrated circuits based on such FETs is the increase in operating frequency, \textit{e.g.} rising the processor operating frequency to higher than nowadays standard of 2 GHz. The evidence of ferroelectric negative capacitance in nanoscale heterostructures has been demonstrated in several experiments \cite{Khan}.
Another possible motivation lays in the field of metamaterials as suggested by Hrabar et al. \cite{Hrabar} The negative capacitance element in their transmission line provided an ultra-broadband (dispersionless) metamaterial. However, this key element was an active electronic circuit of macroscopic scale that, by itself, presents a constraint and limits the size of a device. Therefore, for application purposes it would be of the essence to find a heterostructure on microscopic scale providing the negative electric capacitance with a broadband response.
Very good review paper covering the development in the field related to the ferroelectric negative electric capacitance is published recently by \'{I}\~{n}iguez et al \cite{Zubko}.\\ 
A feasible way to stabilize a working point on a section of polarization vs electric field characteristic with negative electric capacitance was proposed by Rusu et al. \cite{Rusu} based on the Maxwell charge equations \cite{Maxwell} and Tsividis model of the MOS transistor \cite{MOS}. The presented idea is to use a layer of linear dielectric on the top of the layer of ferroelectric where the boundary condition between the two fixes the polarization in the ferroelectric layer to the desired working point depending on the parameters of the junction and materials.
In this work we consider this suggestion and calculate the frequency dependent dielectric response.

\bigskip
\textbf{II. The Model}
\bigskip

\begin{figure}
\centerline{\includegraphics[width=7.0cm]{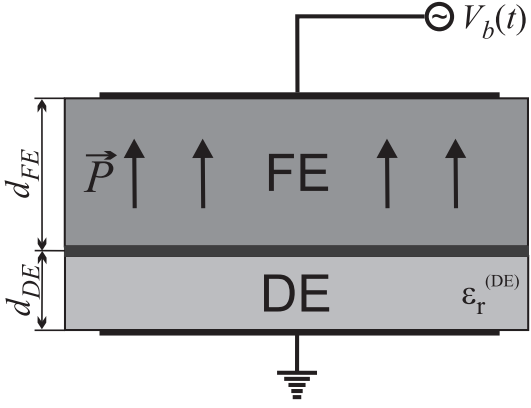}}
\caption{Schematic presentation of layered ferroelectric (FE) - dielectric (DE) junction, with corresponding thicknesses $d_{FE}$ and $d_{DE}$ respectively, biased by time-dependent voltage $V_b(t)$. Here $\vec{P}$ is polarization in the FE layer, while $\epsilon_r^{(DE)}$ is relative dielectric constant of the DE layer. In further modeling we use realistic material parameters. For the FE material parameters are taken for BaTiO$_3$: $a=-2 \cdot 10^7$Vm/C, $b=-35.6 \cdot 10^8$Vm$^5$/C$^3$ and $c=27 \cdot 10^{10}$Vm$^9$/C$^5$ in eq. (\ref{Landau}), with thickness $d_{FE}=55$nm, and polarization switching time $\tau \sim 50 \div 70$ps. \cite{BaTiO3, SwitchingTime} The DE material is Al$_2$O$_3$, with thickness $d_{DE}=7$nm and relative dielectric permittivity $\epsilon_r^{(DE)}=9.1$.}
\label{Figjunction}
\end{figure}

We consider the system schematically shown in fig. \ref{Figjunction}: layer of linear dielectric (DE) on the layer of ferroelectric (FE), biased by time-dependent voltage $V_b(t)$.
%
%
Polarization $\vec{P}$ in ferroelectric exposed to electric field $\vec{E}$ inside it is described by the simplest form of Landau expansion of energy
%
\begin{eqnarray}
U(P)=U_0+\frac{a}{2}P^2+\frac{b}{4}P^4+\frac{c}{6}P^6+...-\vec{E}\cdot\vec{P},
\label{Landau}
\end{eqnarray}
%
where $a,b,c,...$ are coefficients in the expansion depending on the material and external conditions. We limit our consideration to the ferroelectric phase at temperature $T$ below the critical FE transition temperature $T_c$ for which we have coefficient $a \sim T-T_c$ negative, \textit{i.e.} $a=-|a|<0$. From the set of coefficients describing the nonlinear contributions of higher order, we keep only $b$ and $c$ to account for the nonlinearity. Naturally, in a ferroelectric, polarization would get stabilized at the minimum of energy which corresponds to positive electric capacitance.
The time evolution of polarization is determined by Landau-Khalatnikov (L-K) equation \cite{LKtheory} $\rho \dot{\vec{P}}=-\nabla_P U(P)$ which, neglecting the anisotropy of polarization in FE layer and considering it homogeneous along the direction perpendicular to FE-DE interface, reduces to
%
\begin{eqnarray}
\rho \frac{\textrm{d}P}{\textrm{d}t}=-|a|P+bP^3+cP^5-E,
\label{Landau-Kalatnikov}
\end{eqnarray}
%
where $\rho$ is the kinetic coefficient accounting for the "internal resistance" (or "viscosity") causing the delay in polarization switching, here in units of electric resistivity since $dP/dt$ is dimensionally current.

\begin{figure}
\centerline{\includegraphics[width=\columnwidth]{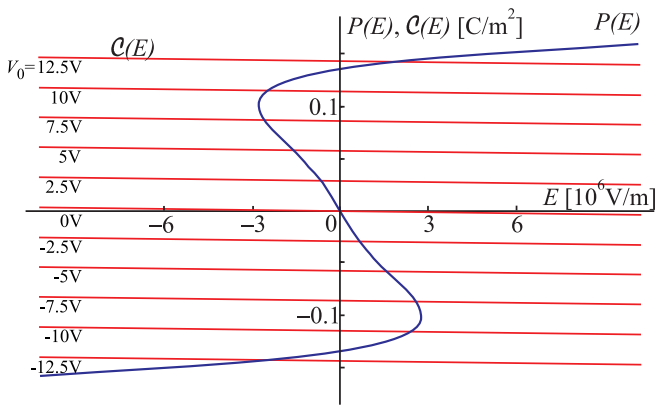}}
\caption{Polarization $P$ dependent on electric field $E$ in the FE layer obtained as a solution of the stationary Landau-Khalatnikov equation eq.(\ref{Landau-Kalatnikov}). FE (BaTiO$_3$) and DE (Al$_2$O$_3$) material parameters are listed in the fig. \ref{Figjunction} caption. Charge lines $\mathcal{C}(E)$ eq.(\ref{charge_line}) are obtained correspondingly within the span of bias voltages with amplitude $V_0$ between -12.5V and 12.5V. It is evident that the negative slope is preserved approximately within the range $|V_0|<8$V, while the linear regime is rather well kept for $|V_0|<2$V.}
\label{FigP(E)}
\end{figure}

\begin{figure}
\centerline{\includegraphics[width=\columnwidth]{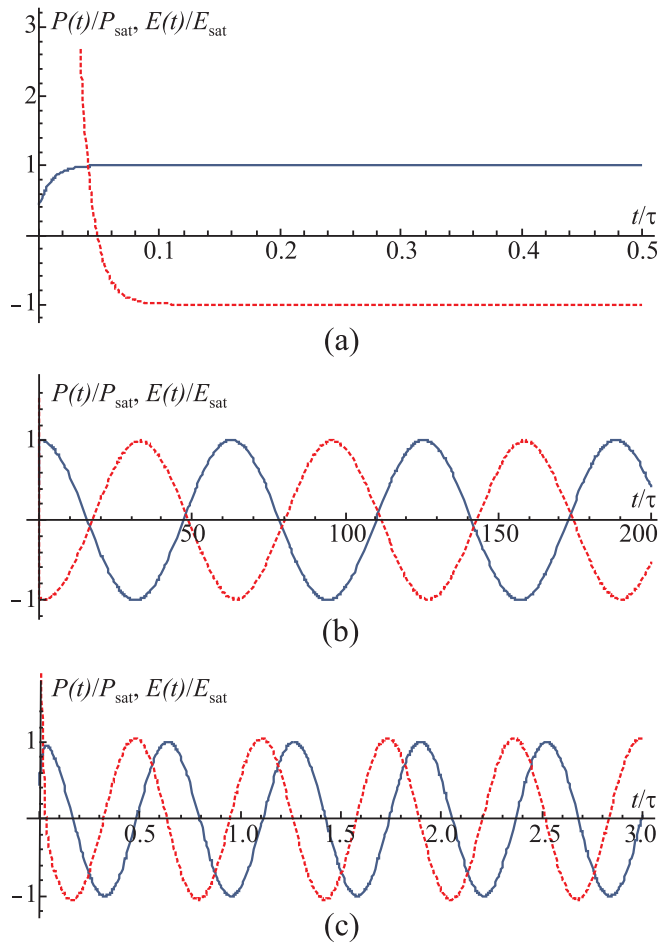}}
\caption{The polarization $P(t)$ (full) and electric field $E(t)$ (dashed) dependent on time $t$ (in units of characteristic polarization switching time $\tau$ - see eq. (\ref{relaxation-time})) in the FE layer obtained as a solution of the eq.(\ref{FE-DE_equation}). Solutions are modeling FE-DE junction in which material parameters are listed in the fig. \ref{Figjunction} caption. Different characteristic frequency regimes (with respect to the time scale $\tau$), imposed by the bias voltage, are shown: (a) for $\omega \tau=0$ \textit{i.e.} stationary bias voltage; (b) for $\omega \tau=0.1$; (c) for $\omega \tau=10$. The frequency dependent difference in phase shift between polarization and electric field is evident. Polarization and electric field in pictures are normalized to their maximal saturation values $P_{sat}$ and $E_{sat}$ respectively.}
\label{FigP(t)}
\end{figure}

The afore mentioned boundary condition at the FE-DE interface \cite{Rusu} ensures continuation of dielectric shift $\vec{D}_{DE}=\vec{D}_{FE}$ in the absence of trapped charge in it. It pins the polarization to the specific (working) point of the $P(E)$ characteristic that can be controlled by the parameters of the junction. Using the auxiliary variables $\vec{E}_{FE}$ and $\vec{E}_{DE}$ for electric fields in ferroelectric and dielectric layer respectively, we can write this condition as
%
\begin{eqnarray}
\epsilon_0\vec{E}_{FE} + \vec{P} = \epsilon_0 \epsilon_r^{(DE)} \vec{E}_{DE},
\label{condition1}
\end{eqnarray}
%
where $\epsilon_0$ is dielectric permittivity of vacuum and $\epsilon_r^{(DE)}$ is relative dielectric permittivity of dielectric layer.
Expressing the auxiliary electric fields in terms of corresponding voltage drops $V_{FE}$ and $V_{DE}$ over each layer of thickness $d_{FE}$ and $d_{DE}$ respectively (\textit{i.e.} $V_{FE}=E_{FE}d_{FE}$, $V_{DE}=E_{DE}d_{DE}$) and taking into account that total voltage drop is $V_b=V_{FE}+V_{DE}$, we transform eq. (\ref{condition1}) into condition that pins the polarization in the FE layer
%
\begin{eqnarray}
P(E_{FE})=\epsilon_0\frac{\epsilon_r^{(DE)}}{d_{DE}}V_b-\epsilon_0 \left(1 + \frac{\epsilon_r^{(DE)}d_{FE}}{d_{DE}} \right) E_{FE}.
\label{condition2}
\end{eqnarray}
%
Writing eq. (\ref{condition2}) we assumed homogeneous field and polarization in layers, as well as their parallel directions as before, so the vectors are omitted. This equation must be solved together with eq. (\ref{Landau-Kalatnikov}).
According to ref. [\cite{Rusu}], the right-hand side of eq. (\ref{condition2}) is so-called "charge line"
%
\begin{eqnarray}
\mathcal{C}(E)=A-BE,
\label{charge_line}
\end{eqnarray}
%
with coefficients
%
\begin{eqnarray}
A(t)=\frac{\epsilon_0\epsilon_r^{(DE)}}{d_{DE}}V_b(t), \nonumber \\
B=\epsilon_0 \left(1 + \frac{\epsilon_r^{(DE)}d_{FE}}{d_{DE}} \right) >0,
\label{coefficients}
\end{eqnarray}
%
that stabilizes solution for polarization on the section with negative electric capacitance, \textit{i.e.} the unstable section where the slope of function $P(E)$ following from the stationary L-K equation (see fig. \ref{FigP(E)}) is negative.
Expressing $E$ from eq. (\ref{charge_line}) at the point $\mathcal{C}(E)=P(E)$ and inserting into eq. (\ref{Landau-Kalatnikov}), we obtain the system of equations that determines polarization and electric field in the ferroelectric layer in FE-DE junction
%
\begin{eqnarray}
\rho \frac{\textrm{d}P(t)}{\textrm{d}t}+\left( \frac{1}{B}-|a| \right)P(t)+bP(t)^3+cP(t)^5 = \frac{A(t)}{B} \nonumber \\
E(t)=\frac{1}{B}\left(A(t)-P(t)\right).
\label{FE-DE_equation}
\end{eqnarray}
%
Comparing the L-K equation with the first equation in system (\ref{FE-DE_equation}), we immediately notice the condition that now stabilizes solution on the section with negative electric capacitance, which is that coefficient associated with $P$ is positive, \textit{i.e.}
%
\begin{eqnarray}
\alpha \equiv \frac{1}{B}-|a|>0,
\label{a-eff}
\end{eqnarray}
%
meaning that the (absolute) slope of $\mathcal{C}(E)$ line is smaller than the slope of $P(E)$ at that section.
This system can be solved numerically, but also analytically within the framework of linear response approximation. The linear response approximation will give us the main insight in the nature of this response. After that we shall present numerical solution of a more realistic nonlinear case and compare the frequency dependence of polarization and phase shift with respect to electric field. It is also possible to formulate the higher orders of nonlinear susceptibilities as suggested by Miga et al [\cite{Miga}], but this goes beyond the scope of this work and its goal, and brings no essential information in that respect, thus it will not be considered.
In fig. \ref{FigP(t)} we see a typical solution of the system (\ref{FE-DE_equation}) with stabilized solution in the sense of eq. (\ref{a-eff}) calculated for FE-DE junction where we took material parameters of FE = BaTiO$_3$, DE = Al$_2$O$_3$: (a) the time independent case ($V_b(t)=V_0$) in which it is clearly visible that polarization $P(t)$ and electric field $E(t)$ in the FE layer are in opposite direction (phase shift between them is $-\pi$); (b), (c) at finite frequency $\omega$ of time dependent bias voltage ($V_b(t)=V_0 cos(\omega t)$) there is a finite frequency-dependent phase shift between $P(t)$ and $E(t)$ which differs from $-\pi$.
%
%
%
%

\bigskip
\textbf{II. The Linear Response}
\bigskip

Within the linear response approximation we treat the system (\ref{FE-DE_equation}) neglecting the higher order terms in $P$ (\textit{i.e.} setting $b=c=0$). The latter is valid for small enough amplitude of bias voltage for which the working point gets positioned to the linear part of $P(E)$ characteristic (around $P(E)=0$). The response function - the electric susceptibility and corresponding dielectric function in frequency domain are calculated as
%
\begin{eqnarray}
\chi(\omega)=\frac{P(\omega)}{\epsilon_0 E(\omega)},\hspace{5mm} \epsilon_r(\omega)=1+\chi(\omega)
\label{susceptibility-general}
\end{eqnarray}
%
respectively.

\bigskip

The stationary susceptibility $\chi_0$ of the FE layer is easily calculated from stationary solution of system (\ref{FE-DE_equation}), \textit{i.e.} stating $\omega=0$, $V_b(t)=V_0$, $A(t)= A_0 \equiv \epsilon_0\epsilon_r^{(DE)}V_0/d_{DE}$ and $\dot{P}=0$, providing
%
\begin{eqnarray}
P=\frac{A_0/B}{\frac{1}{B}-|a|}, \nonumber \\
E=-|a|P.
\label{solution-stationary}
\end{eqnarray}
%
From eq. (\ref{solution-stationary}) there follows
%
\begin{eqnarray}
\chi_0=-\frac{1}{\epsilon_0 |a|}<0.
\label{susc-stationary}
\end{eqnarray}
%
The stationary susceptibility of FE layer is negative providing the corresponding dielectric constant $\epsilon_r^{(0)}=1-(\epsilon_0|a|)^{-1}<1$. This result clearly indicates a stable solution with negative electric capacitance: polarization in the FE layer is directed in opposite direction with respect to the electric field (see eq. (\ref{solution-stationary}), second equation). Here susceptibility is purely real and the phase shift between polarization and electric field is $-\pi$.

\bigskip

\begin{figure}
\centerline{\includegraphics[width=8.0cm]{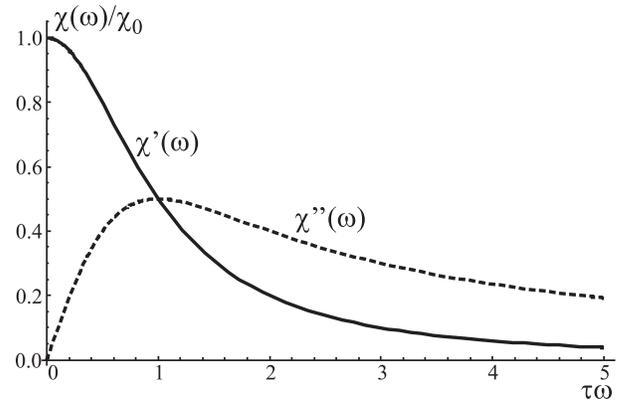}}
\caption{Frequency dependence of susceptibility of ferroelectric layer eq. (\ref{susc-final}) within the linear response theory. $\chi'(\omega)$ real and $\chi"(\omega)$ imaginary part of susceptibility are plotted normalized to $\chi_0<0$. Here $\tau$, defined by eq. (\ref{relaxation-time}), is the polarization switching time characteristic of the FE material.}
\label{FigSusceptibility}
\end{figure}

The frequency dependent electric susceptibility in the FE layer is obtained from the time dependent solution of the system (\ref{FE-DE_equation}) for the simple choice of harmonic bias voltage $V_b(t)=V_0 \cos(\omega t)$ with single frequency $\omega$. Stationary time dependent solution of this system is, for polarization
%
\begin{eqnarray}
P(t)=\frac{\frac{A_0}{B}}{\alpha^2+(\rho\omega)^2} \left( \alpha \cos (\omega t) + (\rho\omega) \sin (\omega t) \right),
\label{polarization-timedependent}
\end{eqnarray}
%
while electric field is then easily found from the second equation of system (\ref{FE-DE_equation}). In solution eq. (\ref{polarization-timedependent}) we neglected the exponentially damped homogeneous solution $P_h(t) \sim \exp{(-\alpha t/\rho)}$ with relaxation time $\rho/\alpha$. Here we also single out the polarization switching time characteristic for sole ferroelectric material,
%
\begin{eqnarray}
\tau = \frac{\rho}{|a|}=\epsilon_0 \rho |\chi_0|.
\label{relaxation-time}
\end{eqnarray}
%
Such switching time highly depends on preparation of the sample. Nowadays samples can achieve switching times of the order of dozens picoseconds. \cite{SwitchingTime}
We can express both $P(t)$ and $E(t)$ in the form
%
\begin{eqnarray}
P(t)=\mathcal{A}_P \cos (\omega t + \phi_P), \nonumber \\
E(t)=\mathcal{A}_E \cos (\omega t + \phi_E),
\label{P-E-representation}
\end{eqnarray}
%
provided
%
\begin{eqnarray}
\mathcal{A}_P = \frac{\frac{A_0}{B\alpha}}{\sqrt{1+ \left( \frac{\rho\omega}{\alpha} \right)^2}}, \nonumber \\
\mathcal{A}_E = \frac{\frac{A_0}{B\alpha}}{\sqrt{1+ \left( \frac{\rho\omega}{\alpha} \right)^2}} \sqrt{|a|^2+(\rho \omega)^2}, \nonumber \\
\tan(\phi_P) = -\frac{\rho \omega}{\alpha}, \nonumber \\
\tan(\phi_E) = \frac{\frac{\rho \omega}{\alpha}}{B\alpha \left( 1+\left( \frac{\rho\omega}{\alpha} \right)^2 \right) -1}.
\label{amplitudes}
\end{eqnarray}
%
Shifting the time origin conveniently, we can write Eqs. (\ref{P-E-representation}) in the more convenient way, \textit{i.e.} $P(t)=\mathcal{A}_P \cos (\omega t + \Delta \phi)$ and $E(t)=\mathcal{A}_E \cos (\omega t)$. There $\Delta \phi = \phi_P-\phi_E$, obtained combining last two expressions in eq. (\ref{amplitudes}) and determined by
%
\begin{eqnarray}
\tan(\Delta \phi) = \frac{\rho \omega}{|a|},
\label{tanDeltaPhi}
\end{eqnarray}
%
is the phase shift between electric field and polarization in the FE layer which, taking into account the correct branch of $\arctan$ function, equals
%
\begin{eqnarray}
\Delta \phi = -\pi + \arctan \left( \tau \omega \right).
\label{DeltaPhi}
\end{eqnarray}
%
$\Delta \phi$ attains the value $-\pi$ in the $\omega=0$ limit and asymptotically approaches the value $-\pi/2$ in the $\tau\omega \gg 1$ limit.
At finite frequency, giving rise to the finite phase shift $\Delta \phi$ different from $-\pi$, the Fourier transforms of $P(t)$ and $E(t)$ in eq. (\ref{susceptibility-general}) yield the complex susceptibility with real and imaginary part
%
\begin{eqnarray}
\chi'(\omega)=\frac{1}{\epsilon_0} \frac{\mathcal{A}_P(\omega)}{\mathcal{A}_E(\omega)} \cos \left( \Delta \phi (\omega) \right), \nonumber \\
\chi"(\omega)=\frac{1}{\epsilon_0} \frac{\mathcal{A}_P(\omega)}{\mathcal{A}_E(\omega)} \sin \left( \Delta \phi (\omega) \right).
\label{susc-real-imaginary}
\end{eqnarray}
%
respectively. After short calculation, expressing $\cos$ and $\sin$ in terms of $\arctan$ function, we get
%
\begin{eqnarray}
\chi'(\omega)= \chi_0 \frac{1}{1+(\tau \omega)^2}, \nonumber \\
\chi"(\omega)=\chi_0 \frac{\tau \omega}{1+(\tau \omega)^2}.
\label{susc-final}
\end{eqnarray}
%
This type of response (see fig. \ref{FigSusceptibility}) is typical so-called relaxation response (Debye) \cite{Dissado}, but we emphasize that $\chi_0$ is negative (see eq. (\ref{susc-stationary})), providing the real part of relative dielectric function $\epsilon_r'(\omega) = 1+\chi'(\omega)<1$! We have Debye-like response of the system characterized by negative electric capacitance, decreasing with frequency, accompanied with dielectric losses appearing at the finite frequencies.
%
%

\bigskip
\textbf{III. Effects of nonlinearity}
\bigskip

\begin{figure}
\centerline{\includegraphics[width=\columnwidth]{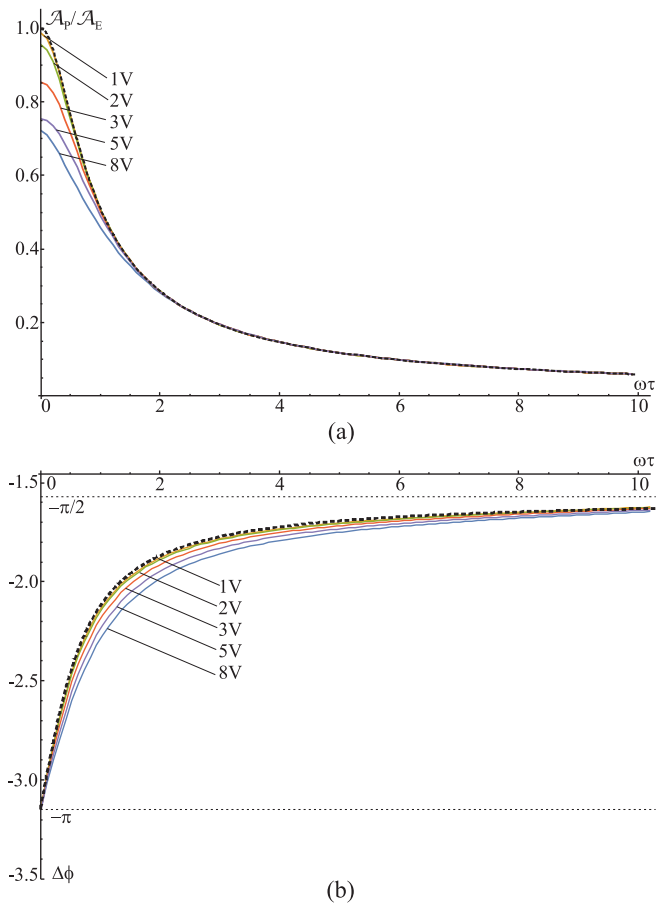}}
\caption{Comparison of the linear model (dashed) and numerical exact solution (full) of eq. (\ref{FE-DE_equation}): (a) ratio of amplitudes of polarization ($\mathcal{A}_P$) and electric field ($\mathcal{A}_E$); (b) phase shift between polarization and electric field in the FE layer depending on frequency $\omega$. FE-DE material parameters used in calculation are listed in the fig. \ref{Figjunction} caption. Ratio of amplitudes is normalized to $|\chi_0|\epsilon_0$. Results are presented for different values of bias voltage amplitude $V_0$ listed in the figure.}
\label{FigNumerical}
\end{figure}

Numerical treatment of eq. (\ref{FE-DE_equation}) opens possibility to address the role of nonlinearity in $P(E)$ dependence. Here we take under consideration the range of $V_0$ covering the negative slope of $P(E)$ characteristic (see fig. \ref{FigP(E)}), thus taking out of consideration branches responsible for hysteretic effects which are of no interest for this analysis anyway. As announced above, we do not consider formulation of higher orders of nonlinear susceptibilities, but rather track the influence of nonlinearity by calculating numerically frequency dependent ratio of amplitudes of polarization and electric field, as well as the phase shift between them, depending on parameter $V_0$ which is proportional to deviation from the linear model (and still kept within the region with negative capacitance). Then we compare numerically calculated ratio $\mathcal{A}_P(\omega) / \epsilon_0 \mathcal{A}_E(\omega)$ with analytical value $|\chi_0|/\sqrt{1+(\omega\tau)^2}$ obtained for linear response from eq. (\ref{amplitudes}), and numerically calculated phase shift $\Delta \phi (\omega)$ with analytical result for linear response determined by eq. (\ref{DeltaPhi}). Results are shown in fig. \ref{FigNumerical}.
%
%
The deviations from linear model are most pronounced in the region of frequency small as compared to the FE characteristic frequency scale $\tau^{-1}$, while for higher frequencies the nonlinear result saturate toward the linear model. As expected, the deviations are more pronounced for a larger value of $V_0$, \textit{i.e.} the response amplitude decreases, but the phase shift changes from $-\pi$ more slowly with increasing $V_0$. However, the nonlinearities do not introduce any fundamental difference in the behaviour of response function with respect to frequency, just mere numerical correction, \textit{e.g.} at worst up to 30 percent in response amplitude at zero frequency.

\bigskip
\textbf{IV. Conclusions}
\bigskip

We considered a layered ferroelectric (FE) - dielectric (DE) junction biased by time dependent harmonic voltage $V_b(t)=V_0 \cos (\omega t)$ with single frequency $\omega$. The electric polarization $P$ in the FE layer is described within the framework of standard Landau expansion and we limit our consideration to regime below the critical temperature of the ferroelectric transition, \textit{i.e.} with finite polarization. Conveniently tailored boundary condition, imposed on the dielectric shift at the FE-DE boundary, pins the polarization in the FE layer and effectively stabilizes a working point of the junction to the region with negative slope of the FE polarization with respect to electric field dependence. This gives rise to a stable solution with negative electric capacitance (otherwise unstable), \textit{i.e.}negative static susceptibility $\chi_0<0$ and relative dielectric constant $\epsilon_r<1$. The time dependent solutions, providing the frequency dependence of susceptibility, are obtained from the Landau-Khalatnikov equation. The frequency dependent linear response yields the typical relaxation (Debye) type of susceptibility, however, preserving its negative sign. At finite frequencies $\omega$ both real and imaginary susceptibility, $\chi'(\omega)=\chi_0/(1+(\tau \omega)^2)$ and $\chi"(\omega)=\chi_0\tau \omega/(1+(\tau \omega)^2)$ respectively, where $\tau$ is polarization switching time characteristic to the ferroelectric material, are finite. They appear as such due to the phase shift between polarization and electric field that attains the value $-\pi$ for $\omega=0$ and saturates to $-\pi/2$ at high frequencies with characteristic scale $\omega_0=2\pi/\tau=2\pi(\epsilon_0\rho|\chi_0|)^{-1}$, where $\rho$ is the kinetic coefficient in units of resistivity, accounting for the "viscosity" causing the delay in polarization switching in the FE material. Thus, $\omega_0$ also represents the upper frequency limit up to which the benefits of negative electrical capacitance are feasible for application. Well tailored material with fast switching time, \textit{e.g.} $\tau$ of the order of 10ps, can provide operational frequencies up to the gigahertz range. The effects of nonlinearities of polarization versus electric field dependence, within the section with negative slope, are addressed numerically, calculated from the Landau-Khalatnikov equation with parameters characterizing the FE material BaTiO$_3$ and the DE material Al$_2$O$_3$. No qualitative deviations from the linear model in the response functions were noticed, just numerical corrections at worse up to 30 percent in the low-frequency limit.

\bigskip

\textbf{\emph{Acknowledgement}}. This work was supported by the Croatian Science Foundation, project IP-2016-06-2289,
and by the QuantiXLie Centre of Excellence, a project cofinanced by the Croatian Government and European Union
through the European Regional Development Fund - the Competitiveness and Cohesion Operational Programme
(Grant KK.01.1.1.01.0004). The authors are grateful to K. Juri\v{s}i\'{c} for the fruitful discussions.


\begin{thebibliography}{99}
\bibitem{Salahuddin} S. Salahuddin and S. Datta, Nano Lett. {\bf 8}, 405 (2008).
\bibitem{Khan} A. I. Khan, D. Bhowmik, P. Yu, S. J. Kim, X. Pan, R. Ramesh, and S. Salahuddin, Appl. Phys. Lett. {\bf 99}, 113501 (2011);
R. Tadros-Morgane, G. Vizdrik, B. Martin, and H. Kliem, J. Appl. Phys. {\bf 109}, 014501 (2011).
\bibitem{Hrabar} S. Hrabar, I. Krois, I. Bonic, and A. Kiricenko, Appl. Phys. Lett. {\bf 99}, 2541103 (2011).
\bibitem{Zubko} J. \'{I}\~{n}iguez, P. Zubko, I. Luk'yanchuk, and A. Cano, Nat. Rev. Mater. {\bf 4}, 243 (2019).
\bibitem{Rusu} A. Rusu, A. Saeidi, and A. M. Ionescu, Nanotechnology {\bf 27}, 115201 (2016).
\bibitem{Maxwell} J. Buck, and W. Hayt, \emph{Engineering Electromagnetics} Vol. {\bf 7} (New York: McGraw-Hill, 2011).
\bibitem{MOS} A. Cano, and D. Jimenez, Appl. Phys. Lett. {\bf 97}, 133509 (2010).
\bibitem{BaTiO3} T. Mistsui, I. Tatsuzaki, E. Nakamura, \emph{An Introduction to the Physics of Ferroelectrics}; Gordon and Breach Science Publishers: London, 1976.
\bibitem{SwitchingTime} Y. Akishige, and Y. Kamishina, \emph{Dc electrical resistivity of reduced hexagonal BaTiO3}, Ferroelectrics, 168:1, 121-125 (1995), DOI: 10.1080/00150199508007854; J. Li, B. Nagaraj, H. Liang, W. Cao, Chi. H. Lee, and R. Ramesh, Appl. Phys. Lett., Vol. {\bf 84}, No. 7, 1174 (2004).
\bibitem{LKtheory} L. D. Landau, I. M. Khalatnikov, \emph{On the anomalous absorption of sound near a second order phase transition point}, Dokl. Akad. Nauk {\bf 1954}, 96, 469-472; T. K. Song, Journal of the Korean Physical Society, Vol. {\bf 46}, No. 1, pp. 5-9 (2005); S. Sivasubramanian, A. Widom, and Y. N. Srivastava, \emph{Physical Kinetics of Ferroelectric Hysteresis}, Ferroelectrics, 300:1, 43-55 (2004), DOI:10.1080/00150190490442173; L.-H. Ong, and K.-H. Chew, in \emph{Ferroelectrics - Characterization and Modeling}, M. Lallart (Ed.), p. 349 (InTech 2011, ISBN: 978-953-307-455-9).
\bibitem{Miga} S. Miga, J. Dec, W. Kleemann, in \emph{Ferroelectrics - Characterization and Modeling}, M. Lallart (Ed.), p. 181 (InTech 2011, ISBN: 978-953-307-455-9)
\bibitem{Dissado} L. Dissado, \emph{Dielectric response}, in \emph{Springer Handbook of Electronic and Photonic Materials - Springer Handbooks}, S. Kasap and  P. Capper (eds), p. 219 (Springer, Cham, 2017, ISBN: 978-3-319-48931-5)
\end{thebibliography}
\end {document}